\documentclass[aps,prd,twocolumn,superscriptaddress,floatfix,nofootinbib,showpacs,amsmath,amssymb,altaffilletter]{revtex4-2}
\pdfoutput=1

\UseRawInputEncoding

\usepackage{amsmath}
\usepackage{amsfonts}
\usepackage{amssymb}
\usepackage{graphicx}
\usepackage{bm}
\usepackage{subfigure}
\usepackage{url}
\usepackage[hyperindex]{hyperref}
\usepackage{color}
\usepackage[ddmmyy,24hr]{datetime}
\usepackage{bigdelim}
\usepackage{booktabs}
\usepackage{dcolumn}
\usepackage{multirow}
\usepackage{subfigure}
\usepackage{physics}
\usepackage{cancel}
\usepackage{stackrel}
\usepackage{paralist}
\usepackage{xspace}
\usepackage{slashed}
\usepackage{cancel}
\usepackage{todonotes}
\usepackage{enumerate}
\usepackage{float}
\usepackage{fullpage}
\usepackage{ulem}
\usepackage{wasysym}
\usepackage{comment}
\usepackage{bbold}
\usepackage{orcidlink}

\usepackage{feynmp}
\newcommand{\nua}[1]{\ensuremath{\rlap{\kern-2.5pt\ensuremath{\overset{\scriptscriptstyle(-)}{\phantom{\nu}}}}{\ensuremath{{\nu}_{#1}}}}}

\usepackage{enumitem}

\begin{document}

\title{Testing Lepton Wave Function Factorization in $^{71}\text{Ge}$ Electron Capture}

\author{M. Cadeddu \orcidlink{0000-0002-3974-1995}}
\email{matteo.cadeddu@ca.infn.it}
\affiliation{Istituto Nazionale di Fisica Nucleare (INFN), Sezione di Cagliari,
	Complesso Universitario di Monserrato - S.P. per Sestu Km 0.700,
	09042 Monserrato (Cagliari), Italy}

\author{N. Cargioli \orcidlink{0000-0002-6515-5850}}
\email{nicola.cargioli@ca.infn.it}
\affiliation{Istituto Nazionale di Fisica Nucleare (INFN), Sezione di Cagliari,
	Complesso Universitario di Monserrato - S.P. per Sestu Km 0.700,
	09042 Monserrato (Cagliari), Italy}
    
\author{M. Cau \orcidlink{0009-0007-9852-6746}}
\email{michelecau55@gmail.com}
\affiliation{Dipartimento di Fisica, Universit\`{a} degli Studi di Cagliari,
	Complesso Universitario di Monserrato - S.P. per Sestu Km 0.700,
	09042 Monserrato (Cagliari), Italy}

\author{F. Dordei \orcidlink{0000-0002-2571-5067}}
\email{francesca.dordei@cern.ch}
\affiliation{Istituto Nazionale di Fisica Nucleare (INFN), Sezione di Cagliari,
	Complesso Universitario di Monserrato - S.P. per Sestu Km 0.700,
	09042 Monserrato (Cagliari), Italy}

\author{L. Ferro \orcidlink{0009-0002-1698-3710}}
\email{luca.ferro@ca.infn.it}
\affiliation{Dipartimento di Fisica, Universit\`{a} degli Studi di Cagliari,
	Complesso Universitario di Monserrato - S.P. per Sestu Km 0.700,
	09042 Monserrato (Cagliari), Italy}
\affiliation{Istituto Nazionale di Fisica Nucleare (INFN), Sezione di Cagliari,
	Complesso Universitario di Monserrato - S.P. per Sestu Km 0.700,
	09042 Monserrato (Cagliari), Italy}

\date{\today}

\begin{abstract}
Electron-capture ratios provide precision tests of atomic wave functions and of possible non-factorization effects in nuclear electron capture. 
By exploiting exact leptonic wave functions, we present a general framework for predicting the electron capture rates of $^{71}\mathrm{Ge}$. Adopting a \textit{non-factorized treatment} of the transition matrix element, we investigate the interplay between the nuclear transition density, which encodes the nuclear structure contribution, and the leptonic wave functions. Using phenomenologically constrained transition densities, we quantify the impact of non-factorization effects on the capture rates. Finally, we compare our theoretical predictions for the $L/K$, $M/K$ and $M/L$ electron capture ratios with the current experimental world averages, providing an up-to-date assessment of the theoretical and experimental status of $^{71}\mathrm{Ge}$ electron capture.
\end{abstract}

\maketitle

\section{Introduction}
Germanium detectors represent the frontier for rare-event searches as they provide excellent energy resolution, low background and efficient scalability. Thus, in the last decades, germanium has been extensively employed in neutrino and dark matter detectors, with the aim of lowering the thresholds. One intrinsic problem of high-purity germanium detectors stems from the activation due to the interactions of neutrons in the target material, which produce the isotope $^{71}\text{Ge}$. The latter is unstable and undergoes electron capture (EC), i.e. the capture of an atomic electron by a nuclear proton which effectively results in the production of a neutron and the emission of a neutrino,  with a half-life of $t_{1/2}^{\rm exp}\,[\text{d}]=11.465(3)$~\cite{Collar:2023yew,PhysRevC.31.666,newGelifetime,PhysRevC.109.055501}. It represents a background source for rare-event searches but also a useful resource to calibrate the detector response.
Therefore, a robust theoretical description of such a capture process is critical for correctly interpreting the experimental measurements. \\
The process under investigation in this work is the electron capture of ${}^{71}\text{Ge}$ to the ground state of ${}^{71}\text{Ga}$ 
\begin{equation}
e^-_{b} + {}^{71}\text{Ge} \rightarrow {}^{71}\text{Ga} + \nu_e \,.
    \label{eq:process}
\end{equation}
This transition is classified as an allowed Gamow-Teller (GT) EC, with the selection rules $\Delta J^{\rm P} = 1^+$, where $J$ is the total angular momentum, and P represents the parity.
\\
The total energy released for this reaction is determined by the atomic mass difference between the initial and final states in their ground states, namely $Q = 232.47\pm0.09 \text{ keV}$ \cite{Qvalue}. For a capture from an \mbox{$x$-orbital}, the emitted neutrino energy $E_{\nu,x}$ is given by 
\begin{equation}
E_{\nu,x} = Q  -R_\gamma- R_x\,,
    \label{eq:neutrino_energy}
\end{equation}
where $R_x$ is the so-called atomic relaxation energy after the capture from the $x$-orbital and $R_\gamma$ the energy of the nuclear excited state. Given that in this work we consider ECs to the ground state of $^{71}\text{Ga}$ only, no nuclear de-excitation occurs and consequently $R_\gamma = 0$. In this scenario, Eq.~\eqref{eq:neutrino_energy} reduces to $E_{\nu,x} = Q - R_x$. 
\\

The standard theory of orbital EC is reviewed in Ref.~\cite{Bambynek}, where the leptonic wave functions are expanded at leading order
inside the nuclear volume and factorized from the nuclear matrix element. This
works well when the leptonic wave functions vary slowly within the nucleus, or when
their radial dependence can be trivially absorbed into simple shell-dependent Coulomb
amplitudes. 

However, recently it has been discussed that this factorization may not be sufficient to properly interpret certain experimental observations, and thus should be reassessed~\cite{Cadeddu:2025ueh}. Having in hand the exact Dirac-Hartree-Fock-Slater (DHFS) radial electron wave functions obtained by the RADIAL package~\cite{radial}, 
we determine the full amplitudes for the transition, before applying the canonical leading order approximation. 
\\ 
The capture rate for an electron in the atomic shell $x$ is defined within the full formalism as an incoherent sum over the neutrino partial waves $\kappa_\nu$~\cite{n6d8-cmcg}
\begin{equation}
    \lambda_{x} = \frac{G_F^2 V_{ud}^2}{2\pi} E_{\nu,x}^2 \frac{g_A^2}{2J_i+1} \sum_{\kappa_\nu} |\mathcal{M}(\kappa_e, \kappa_\nu)|^2 \mathbf{B}_x\mathbf{S}_x \,,
    \label{eq:EC_rate}
\end{equation}
where $G_F$ denotes the Fermi constant, $V_{ud}$ is the CKM matrix element~\cite{ParticleDataGroup:2024cfk} and $g_A=1.2764$ is the axial-vector coupling constant~\cite{PERKEO3}. Here, $J_i = 1/2$ is the nuclear angular momentum of the initial $^{71}\text{Ge}$ state, $\mathbf{B}_x$ is the atomic correction factor introduced to account for the exchange and overlap corrections~\cite{Bambynek,Elliott:2023xkb,PhysRevLett.9.500,Vatai:1970ldf,PhysRev.131.1756,Bahcall:1963zza,BAHCALL1965267,Mougeot:2018whu,Sevestrean:2023qzm}, and $\mathbf{S}_x$ incorporates the atomic corrections related to shake-up and shake-off effects~\cite{Mougeot:2018whu,Sevestrean:2023qzm}.
\\
The total transition amplitude $\mathcal{M}$ entering Eq.~\eqref{eq:EC_rate} 
represents the exact spatial overlap of the initial and final states over the weak interaction operator. Considering only the dominant contribution, the amplitude reads
\begin{align}
\mathcal{M}(\kappa_e, \kappa_\nu) =& \int_0^\infty dr \, r^2 \rho_{1,1}^{TD}(r) \Big[ g_{\kappa_\nu} G^x_{\kappa_e} S_{101}(\kappa_\nu, \kappa_e) \nonumber\\
&- f_{\kappa_\nu} F^x_{\kappa_e} S_{101}(-\kappa_\nu, -\kappa_e) \Big]\, .
    \label{eq:weak_amplitude}
\end{align}
Within this integral, $\rho^{TD}_{JS}(r)\equiv\rho_{1,1}^{TD}(r)$ represents the nuclear transition density, which maps the spatial overlap of the initial and final nuclear wave functions through the weak interaction operator for the considered transition. The latter can be defined as 
\begin{align}
 \rho^{TD}_{1,1}(r)\!=\!\langle f ||\int d\Omega\,  \{Y_0\otimes \sigma_1\}_1 \tau^-|| i \rangle\,,
    \label{eq:BGT}
\end{align}
where $\sigma_1\tau^-$ is the local one-body Gamow-Teller operator. For compactness, both the sum over all nucleons and the radial localization of the one-body operator at the coordinate $r$ are left implicit.
$Y_0$ is the relevant spherical harmonic component and the integral runs over the solid angle $\Omega$ and $i$($f$) represents the initial (final) nuclear state.
\\
The bound electron state for an $x-$orbital is described by the four-component relativistic Dirac spinor
\begin{equation}
    \Psi^x_{\kappa_e}(\mathbf{r}) = \begin{pmatrix} G^x_{\kappa_e}(r) \chi_{\kappa_e}^{\mu_e} \\ i F^x_{\kappa_e}(r) \chi_{-\kappa_e}^{\mu_e} \end{pmatrix}\, ,
    \label{eq:electron_spinor}
\end{equation}
where $\chi_{\kappa_e}^{\mu_e}$ represent the standard spin-angular functions, and $\kappa_e$ is the relativistic Dirac quantum number. The terms $G^x_{\kappa_e}(r)$ and $F^x_{\kappa_e}(r)$ denote the large and small components of the bound electron's radial wave function for the $x-$orbital, respectively. They are obtained by solving the coupled radial DHFS equations for the exact atomic central potential $V(r)$
\begin{equation}
    \begin{aligned}
        \left(\frac{d}{dr}\! +\! \frac{\kappa_e+1}{r}\right)\!G^x_{\kappa_e}(r)\! -\! (E^x_e - V(r) + m_e)F^x_{\kappa_e}(r) &= 0 \, ,\\
        \left(\frac{d}{dr}\! -\! \frac{\kappa_e-1}{r}\right)\!F^x_{\kappa_e}(r)\! +\! (E^x_e - V(r) - m_e)G^x_{\kappa_e}(r) &= 0\, .
    \end{aligned}
    \label{eq:dirac_radial}
\end{equation}
Here, $E^x_e$ is the total electron energy and $m_e$ is the electron mass. The term $V(r)$ is the effective atomic potential, expressed as \mbox{$V(r) = V_{\text{nuc}}(r) + V_{\text{el}}(r) + V_{\text{ex}}(r)$} and $V_{\text{nuc}}(r)$ is the potential generated by the finite-size nucleus, typically modeled with a two-parameter Fermi (2pF) charge distribution. The term $V_{\text{el}}(r)$ accounts for the electrostatic interaction energy with the entire atomic electron cloud, while $V_{\text{ex}}(r)$ is the exchange potential, defined according to the Slater approximation~\cite{Cadeddu:2025pue}.\\
The emitted neutrino is expanded into spherical partial waves, with its large and small components described by exact spherical Bessel functions
\begin{align}
    &g_{\kappa_\nu}(E_{\nu,x}\,r) = j_{l_{\kappa_\nu}}(E_{\nu,x}\,r)\,,\nonumber\\
    &f_{\kappa_\nu}(E_{\nu,x}\,r) = \text{sgn}(\kappa_\nu) j_{\bar{l}_{\kappa_\nu}}(E_{\nu,x}\,r)\,.
    \label{eq:neutrino_wf}
\end{align} 
The remaining factors $S_{SLJ}$ are the geometric angular momentum coupling coefficients. They are analytically defined as \cite{n6d8-cmcg}
\begin{align}
&S_{SLJ}(\kappa_e, \kappa_\nu)\!=\!
(l_{\kappa_\nu} 0 \, l_{\kappa_e} 0 | L 0)\begin{Bmatrix} l_{\kappa_e} & 1/2 & j_{\kappa_e} \\ l_{\kappa_\nu} & 1/2 & j_{\kappa_\nu} \\ L & S & J \end{Bmatrix}\, \\\nonumber
&\times\!\sqrt{2(2j_{\kappa_\nu}+1\!)(2j_{\kappa_e}+1\!)(2l_{\kappa_\nu}+1\!)(2l_{\kappa_e}+1\!)(2S+1\!)}\, ,
\label{eq:S_coeff}
\end{align}
where $S=1$, $L=0$, and $J=1$ for this specific transition, and the brackets $\{\}$ indicate the 9-j symbol.
\\
In these expressions, $\kappa$ represents the relativistic Dirac quantum number, which uniquely determines both the total angular momentum $j_\kappa$ and the orbital angular momentum $l_\kappa$ of the leptonic state \cite{Bambynek}. These quantities are defined according to the standard mapping
\begin{equation}
    j_\kappa = |\kappa| - \frac{1}{2}, \quad l_\kappa = \begin{cases} \kappa & \text{if } \kappa > 0\, , \\ -\kappa-1 & \text{if } \kappa < 0\, . \end{cases}
    \label{eq:quantum_numbers}
\end{equation}
Furthermore, the orbital angular momentum associated with the small component of the wave function is denoted as $\bar{l}_\kappa \equiv l_{-\kappa}$ \cite{Bambynek}. This notation ensures a consistent description of the parity and angular properties of the electron and neutrino spinors within the multipole expansion.\\
A summary of the allowed quantum numbers for the EC process of $^{71}\text{Ge}$ are reported in Tab.~\ref{tab:kvalues}.

\begin{table}[htbp]
    \centering
    \resizebox{0.5\columnwidth}{!}{
    \renewcommand{\arraystretch}{0.4}
    \begin{tabular}{c|c|c|c}
        \hline\hline
        \textbf{Shell} & $k_e$ & $(l_\nu,j_\nu)$ & $k_\nu$ \\[+0.5ex] \hline
        \multirow{2}{*}{$s_{1/2}$} &  \multirow{2}{*}{-1}& (0,1/2) &  $\mathbf{-1}$ \\
        & & (2,3/2) & 2\\[+1.ex]
        \multirow{2}{*}{$p_{1/2}$} &  \multirow{2}{*}{+1}& (1,1/2) &  $\mathbf{+1}$ \\
        & & (1,3/2) & -2\\ [+1.ex]
        \multirow{3}{*}{$p_{3/2}$} &  \multirow{3}{*}{-2}& (1,1/2) &  $\mathbf{+1}$ \\
        & & (1,3/2) & $\mathbf{-2}$\\ 
        & & (3,5/2) & +3\\ [+1.ex]
        \multirow{3}{*}{$d_{3/2}$} &  \multirow{3}{*}{+2}& (0,1/2) &  $\mathbf{-1}$ \\
        & & (2,3/2) & $\mathbf{+2}$\\ 
        & & (2,5/2) & -3\\ [+1.ex]
        \multirow{3}{*}{$d_{5/2}$} &  \multirow{3}{*}{-3}& (2,3/2) &  $\mathbf{+2}$ \\
        & & (2,5/2) & $\mathbf{-3}$\\ 
        & & (4,7/2) & +4\\[+0.5ex] 
        \hline\hline
    \end{tabular}}
    \caption{Allowed quantum numbers for the EC from the $K,\, L$ and $M$ shells of $^{71}\text{Ge}$. In bold, the dominant contributions for each orbital.}
    \label{tab:kvalues}
\end{table}

\section{Explicit radial amplitudes}
We can now derive the explicit radial amplitudes for the specific atomic shells of interest starting from the general expression in Eq.\,\eqref{eq:weak_amplitude} and evaluating the specific angular coefficients $S_{101}(k_e,k_\nu)$. For each atomic orbital, indicated by the value of $\kappa_e$, we consider only the dominant contribution in the neutrino channel expansion corresponding to the values of $\kappa_\nu$ reported in bold in Tab.~\ref{tab:kvalues}. 
\\
For $s_{1/2}$-type orbitals, such as for the $K,\, L_1$ and $M_1$ shells, we consider $(\kappa_e,\kappa_\nu) =( -1,-1)$, while for $p_{1/2}$-type orbitals, such as for the $L_2$ and $M_2$ shells, the dominant contribution corresponds to $(\kappa_e,\kappa_\nu) =( +1,+1)$. Substituting the corresponding quantum numbers into Eq.\,\eqref{eq:weak_amplitude}, the explicit amplitudes read
\begin{align}
\mathcal{M}_{s_{1/2}} \simeq\sqrt{2}\int_0^\infty &dr \, r^2 \rho_{1,1}^{TD}(r) \Big[ G^x_{-1}(r) j_{0}(E_{\nu,x} r) \nonumber\\
&+\dfrac{1}{3}F^x_{-1}(r) j_{1}(E_{\nu,x} r) \Big]\,,
\label{eq:amplitude_K}
\end{align}
\begin{align}
\mathcal{M}_{p_{1/2}} \simeq\sqrt{2}\int_0^\infty &dr \, r^2 \rho_{1,1}^{TD}(r) \Big[ \dfrac{1}{3}G^x_{1}(r) j_{1}(E_{\nu,x} r)\nonumber\\
&-F^x_{1}(r) j_{0}(E_{\nu,x} r) \Big]\,.
\label{eq:amplitude_L2}
\end{align}
for $s_{1/2}$-type and $p_{1/2}$-type orbitals, respectively.\\
For $p_{3/2}$-type orbitals, such as for the $L_3$ and $M_3$ shells, two dominant neutrino channels are found, namely corresponding to $\kappa_\nu = +1,-2$. Therefore two amplitudes have to be determined and evaluating the proper angular contributions, the explicit amplitudes are given by
\begin{align}
&\mathcal{M}^{k_\nu=+1}_{p_{3/2}} \simeq\frac{4}{3}\int_0^\infty dr \, r^2 \rho_{1,1}^{TD}(r) \Big[ G^x_{-2}(r) j_{1}(E_{\nu,x} r) \Big]\,,\\
&\mathcal{M}^{k_\nu=-2}_{p_{3/2}} \simeq -2\sqrt{5}\int_0^\infty dr \, r^2 \rho_{1,1}^{TD}(r) \Big[ \frac{1}{3}G^x_{-2}(r) j_{1}(E_{\nu,x} r)\nonumber\\
&\hspace{1.5 cm}+\frac{1}{5}F^x_{-2}(r) j_{2}(E_{\nu,x} r) \Big]\,,
\label{eq:amplitude_L3}
\end{align}
from which it can be seen that the dominant contribution scales with the $j_1$ Bessel function, instead of $j_0$ term present inside Eq.~\eqref{eq:amplitude_K} for $s_{1/2}$-type and in Eq.~\eqref{eq:amplitude_L2} for $p_{1/2}$-type orbitals.\\
Similar considerations apply also to $d_{3/2}$-type ($M_4$ shell) and $d_{5/2}$ ($M_5$ shell) orbitals. The former can be obtained considering $\kappa_e = +2$ and $\kappa_\nu = -1,+2$, while the latter corresponds to $\kappa_e = -3$ and $\kappa_\nu =+2,-3$, as reported in Tab.~\ref{tab:kvalues}.
The explicit amplitudes can then be derived from Eq.\,\eqref{eq:weak_amplitude}, and, in the case of $d_{3/2}$-orbitals, they read
\begin{align}
&\mathcal{M}^{k_\nu=-1}_{d_{3/2}} \simeq\frac{4}{3}\int_0^\infty dr \, r^2 \rho_{1,1}^{TD}(r) \Big[F^x_{2}(r) j_{1}(E_{\nu,x} r) \Big]\,\\
&\mathcal{M}^{k_\nu=+2}_{d_{3/2}} \simeq -2\sqrt{5}\int_0^\infty dr \, r^2 \rho_{1,1}^{TD}(r) \Big[ \frac{1}{5}G^x_{2}(r) j_{2}(E_{\nu,x} r)\nonumber\\
&\hspace{1.5 cm}-\frac{1}{3}F^x_{2}(r) j_{1}(E_{\nu,x} r) \Big]\,.
\label{eq:amplitude_M4}
\end{align}
while, for $d_{5/2}$-orbitals, they become
\begin{align}
&\mathcal{M}^{k_\nu=+2}_{d_{5/2}} \simeq-\frac{4}{\sqrt{5}}\int_0^\infty dr \, r^2 \rho_{1,1}^{TD}(r) \Big[ G^x_{-3}(r) j_{2}(E_{\nu,x} r)\Big]\,,\\
&\mathcal{M}^{k_\nu=-3}_{d_{5/2}} \simeq\sqrt{\frac{14}{5}}\int_0^\infty dr \, r^2 \rho_{1,1}^{TD}(r) \Big[ G^x_{-3}(r) j_{2}(E_{\nu,x} r) \nonumber\\
&\hspace{1.5 cm}+\frac{5}{7}F^x_{-3}(r) j_{3}(E_{\nu,x} r) \Big]\,.
\label{eq:amplitude_M5}
\end{align}
The main orbitals contributing to the EC rates are those which are proportional to the zero-order Bessel function $j_0$, as in proximity of the origin, higher order Bessel functions result in large suppression. By looking at the obtained amplitudes, it is thus evident that the larger contributions will arise from the $K,\, L_1,\, L_2,\, M_1\,\text{and}\,M_2$ shells. \\
To properly evaluate the numerical value of the capture rates, it is necessary to know the exact form of the transition density $\rho_{1,1}^{TD}(r)$.
Unfortunately, the evaluation of such a quantity  
is rather complex, as it requires the knowledge of the nuclear wave function of the initial and final nuclear states. This is computationally challenging and has been only achieved for light nuclei~\cite{saori-pastore}, and not for mid-size nuclei such as germanium. \\
An alternative approach could consist in parametrizing the transition density with phenomenological functions built starting from basic microscopic descriptions of nuclear dynamics, such as shell model nuclear wave functions or harmonic oscillator approximations of the nucleus. Subsequently, the obtained function could be fitted to the data in order to extract information about its size and shape.\\
The latter approach was employed in a previous work~\cite{Cadeddu:2025ueh} in which a reassessment of the long standing gallium anomaly was performed.
In particular, the gallium anomaly consists of a significant ($\sim 20\%$) deficit~\cite{Cadeddu:2025pue} in the measurement of the cross section for the neutrino capture on $^{71}\text{Ga}$
\begin{equation}
\nu_e + {}^{71}\text{Ga}\rightarrow  {}^{71}\text{Ge}+e^- \,,
    \label{eq:neucapt}
\end{equation}
with respect to its theoretical expectation.
Given that the neutrino capture process in Eq.~\eqref{eq:neucapt} is approximately the time reversal\footnote{Indeed, the two processes probe inverse nuclear transitions and are governed by the same leading Gamow-Teller matrix element in the factorized approximation. In the non-factorized treatment, however, their different leptonic kernels make their relation dependent on the radial transition density.} of the electron capture process on $^{71}\text{Ge}$ in Eq.~\eqref{eq:process}, the measurement of the germanium half-life allows one to infer the expected neutrino capture cross section.
In particular, in Ref.~\cite{Cadeddu:2025ueh} it was shown that, a properly modeled transition density was able to reproduce the measured value of the germanium half-life while simultaneously reproducing the observed value of the neutrino capture cross section on $^{71}\text{Ga}$.
This solution could not be obtained in the standard factorized scheme as in the latter treatment the dependence on nuclear structure cancels out when combining the two different processes. Instead, the introduction of the transition density in the complete non-factorized approach, allows for possible interference in one process while preserving the other. \\
In Ref.~\cite{Cadeddu:2025ueh}, the electron capture process on $^{71}\text{Ge}$ description
treated explicitly the $K$-shell contribution to ground-state capture, while the other orbitals were accounted for by introducing the so-called capture ratios, i.e. the ratio between the capture rate from two different shells. The latter have been measured and thus were taken as data-driven rescaling factors. However, it is important to reconsider the theoretical description of such capture ratios to assess their dependence on the transition density. 
\begin{figure}[t!]
    \centering
    \includegraphics[width=0.9\linewidth]{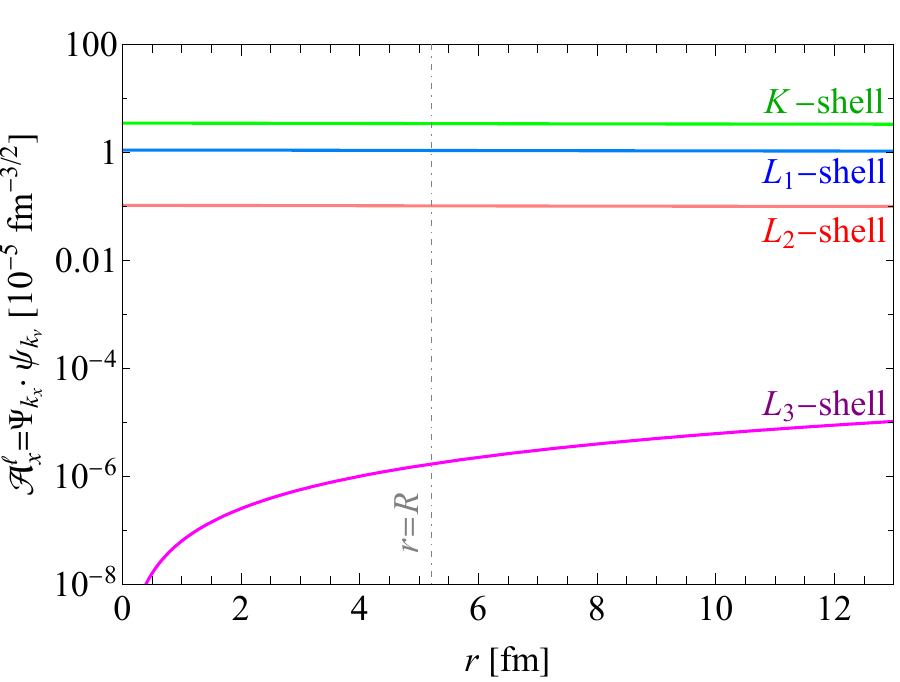}
    \caption{Leptonic amplitudes, defined as product of electron and neutrino wave functions, relevant for the calculation of the EC rate for the $K$, $L_1$, $L_2$ and $L_3$ shells of $e_b+{}^{71}\mathrm{Ge}\to{}^{71}\mathrm{Ga}+\nu_e$ transition as a function of the radial distance from the origin. The dot dashed gray vertical line represents the nuclear radius $R$ of ${}^{71}\mathrm{Ge}$.}
    \label{fig:leptAmp}
\end{figure}

\section{About the validity of the Factorization Scheme}
The starting point is the leptonic kernels entering the various amplitudes, $\mathcal{A}_x^\ell$. These kernels are obtained as the product of the electron and neutrino wave functions. Since the terms proportional to $j_1(E_{\nu,x}r)$ (or higher order Bessel functions) are order of magnitude smaller than those proportional to $j_0(E_{\nu,x}r)$, we can restrict the discussion only to the dominant contribution to each orbital, which are shown in Fig.~\ref{fig:leptAmp}.\\
The dominant contribution to $s_{1/2}$ orbitals is almost flat, as the $G^x_{-1}(r)$ wave function is approximately flat near the origin, and the $j_0(E_{\nu,x}r)$ neutrino contribution is constant. However, for other shells, the product of electron and neutrino wave functions exhibits an $r^{\alpha_x}$ radial dependence, with $\alpha_x$ being characteristic of the specific atomic shell under investigation.\\
In this sense, it is useful to define the effective nuclear transition probed by EC from an $x-$shell as
\begin{align}
 B^{\text{eff},x}_{1,1}(1^+)\!=\!\! \frac{g_A^2}{2J_i+1}\! \left| \int dr\, r^2\left(\frac{r}{R}\right)^{\alpha_x}\! \rho_{1,1}^{TD}(r)  \right|^2,
    \label{eq:BGT}
\end{align}
with $R$ being the reference nuclear radius, introduced to make the radial weight $(r/R)^{\alpha_x}$ dimensionless. The latter  represents the residual radial dependence intrinsic of the leptonic amplitude $\mathcal{A}^\ell_x$ for the specific $x-$shell contribution.
The quantity $B^{\text{eff},x}_{1,1}$ should not be identified with the intrinsic nuclear $B(\text{GT})$, but represents a shell-dependent, kernel-weighted radial moment of the transition density. Consequently, different atomic shells can probe different radial moments of the same nuclear transition density.
\\
From here, it is evident that the effective strength may differ for EC happening from different atomic shells, depending on the dominant $r^{\alpha_x}-$dependence intrinsic of that specific shell. As visible in Fig.~\ref{fig:leptAmp}, the $K,\, L_1$ and $L_2$ shells show a practically flat leptonic amplitude, and thus are expected to share the same effective strength at LO. The same reasoning would apply also to the non-shown $M_1$ and $M_2$ shells, having all $\alpha_{K,L_1,L_2,M_1,M_2}\sim 0$ (i.e. approximately flat wave functions within the nuclear environment). On the contrary, the $L_3$ and $M_3$ leptonic amplitudes have an intrinsic $\mathcal{A}^\ell_{L_3,M_3}\propto r^2$ radial dependence $(\alpha_{L_3,M_3}\sim2)$ which makes their corresponding effective strength different from the one for the other shells.
\\
The transition density in Eq.~\eqref{eq:BGT} is the same for all the shells, as it only depends on the initial and final nuclear states and on the nuclear transition operator. The intrinsic $r^{\alpha_x}$ radial dependence of leptonic wave functions is accounted for by employing the appropriate unexpanded numerical DHFS electron wave functions and the appropriate Bessel-$j$ wave functions describing the neutrino inside the $x-$shell transition amplitude.\\
From this general discussion, it is possible to understand that the eventual interference between the radial profile of the transition density with that of the leptonic kernel is expected to be very similar among those transitions which share the same dominant radial dependence, while it would differ for the others. The size of such a difference would depend explicitly on the actual transition density dominating the process.
In this sense, the $K,\, L_1\,, L_2\,, M_1\, \text{and}\, M_2$ shells are expected to receive a similar nuclear contribution, whereas shells such as the $L_3$ can receive a different contribution.\\
The main advantage of considering electron capture ratios lies in the cancellation of the nuclear matrix element between numerator and denominator after the application of the factorization scheme. However, for the aforementioned reasons, one would expect this cancellation to hold reasonably well when considering ratios of similarly $r^{\alpha_x}$-dependent shells. Therefore, the ratios of $L_1,\, L_2,\, M_1\, \text{and}\, M_2$ to the $K$ shell are expected to be practically independent of the effect of nuclear structure, while the $L_3$ to $K$ ratio may be significantly dependent on the transition density. Luckily, given that the leptonic amplitude of the $L_3$ shells is highly suppressed (about six orders of magnitudes), as clearly visible in Fig.~\ref{fig:leptAmp}, it is unlikely that any possible effect of the transition density could be visible in the available experiments. For these reasons, the prediction of the observable capture ratios of $^{71}\text{Ge}$ obtained in the factorized scheme appears to be sufficiently robust for interpreting the available measurements.

\section{Numerical Results}
In the factorized allowed-LO approximation, we retain only the dominant term in the neutrino wave function expansion, therefore considering the terms at the order $j_0(E_{\nu,x}r)$. Doing so, only the $s_{1/2}$ and the $p_{1/2}$ transitions contribute.
Applying such a reduction, we can define the $L$ to $K$ shell capture ratio as
\begin{equation}
    L/K=\sum_{x=1}^{2} L_x/K\, ,
\end{equation}
where the term $L_3$ has been neglected, as it depends on $j_1$ or higher neutrino contributions.\\
A similar discussion applies to the $M$ to $K$ capture ratio, defined as
\begin{equation}
    M/K=\sum_{x=1}^{2} M_x/K\, ,
\end{equation}
where, once again, only $M_1$ and $M_2$ contribute at LO.\\
For these transitions, the nuclear part of the interaction effectively cancels out when evaluating the capture ratios. Thus, we can provide an operative definition for the EC ratios, which reads
\begin{align}
    \dfrac{L}{K}&\,\stackrel{\rm LO}{=}\, \dfrac{L_1}{K}\!+\!\dfrac{L_2}{K}\!=\!\Bigg[\dfrac{E_{\nu,L_1}}{E_{\nu,K}} \dfrac{\langle G^{L_1}_{-1}(r)j_0(E_{\nu,L_1}r)\rangle }{\langle G^K_{-1}(r)j_0(E_{\nu,K}r)\rangle}\Bigg]^2\frac{\mathbf{B}_{L_1}\,\mathbf{S}_{L_1}}{\mathbf{B}_K\,\mathbf{S}_{K}}\nonumber\\
    &\hspace{1 cm}+\Bigg[\dfrac{E_{\nu,L_2}}{E_{\nu,K}} \dfrac{\langle F^{L_2}_{1}(r)j_0(E_{\nu,L_2}r)\rangle}{\langle G^K_{-1}(r)j_0(E_{\nu,K}r)\rangle}\Bigg]^2\frac{\mathbf{B}_{L_2}\,\mathbf{S}_{L_2}}{\mathbf{B}_K\,\mathbf{S}_{K}}\, ,\label{LtoKatLO}
\end{align}

where a superscript to the electron wave functions has been inserted to distinguish between the same wave functions calculated for a different electron energy. Moreover, we perform an averaging of the leptonic wave functions over the nuclear charge densities, similarly to Ref.~\cite{Cadeddu:2025pue,Elliott:2023xkb}.\\
The $M/K$ ratio at LO has the exact same definition, with $L$ replaced by $M$ inside Eq.~(\ref{LtoKatLO}). The same wave functions enter the definition but for the different electron and neutrino energies. 
These expressions agree with those reported in the literature, for example, see Refs.~\cite{Bambynek,SCHONFELD19981353}.
Combining the definitions for the $L/K$ and $M/K$ ratios, it is also possible to predict the $M/L$ capture ratio.
\begin{table}[htbp]
    \centering
    \resizebox{0.75\columnwidth}{!}{
    \renewcommand{\arraystretch}{1.1}
    \begin{tabular}{c|c|c}
        \hline\hline
        \hspace{0.4 cm}\textbf{$x-$shell}\hspace{0.35 cm}\,& \hspace{0.5 cm}\textbf{$\text{B}_x$}\hspace{0.45 cm}\, & \hspace{0.5 cm}\textbf{$\text{S}_x$}\hspace{0.45 cm}\,  \\ \hline
        $K$ & $0.9901$ & $1.0013$   \\\hline
        $L_1$ & $1.0813$  & $1.0067$   \\
        $L_2$ & $0.9806$ & $1.0055$ \\
        $L_3$ & $0.9578$ & $1.0055$\\\hline
        $M_1$ & $1.217$  & $1.0157$ \\
        $M_2$ & $1.103$ & $1.0156$ \\
        $M_3$ & $1.018$&$1.0156$\\
        $M_4$ & $0.977$&$1.0166$\\
        $M_5$ & $0.977$& $1.0167$
        \\\hline\hline
    \end{tabular}}
    \caption{Exchange and overlap, $\textbf{B}_x$, and shake-up and shake-off, $\textbf{S}_x$, atomic corrections applied in the calculation of the capture ratios, calculated following Refs.~\cite{Sevestrean:2023qzm}.}
    \label{tab:correction_factors}
\end{table}

In the determination of these capture fractions, we evaluated the exchange and overlap corrections, $\mathbf{B}_x$, as well as the shake-up and shake-off effects, $\mathbf{S}_x$, following closely the procedure described in Ref.~\cite{Sevestrean:2023qzm}. By considering our electron wave functions, we obtain the values reported in Tab.~\ref{tab:correction_factors} for the various shells.\\
Finally, the numerical results for the $L/K$, $M/K$ and $M/L$ 
EC ratios at LO in the factorized scheme are reported in Tab.~\ref{tab:theoretical_fractions}, with an associated uncertainty obtained by propagating the error on the $Q-$value together with the uncertainty associated to our electron wave functions, estimated by varying the exchange potential as described in Ref.~\cite{Cadeddu:2025pue}.
\begin{table}[h!]
    \centering
    \resizebox{\columnwidth}{!}{
    \renewcommand{\arraystretch}{1.5}
    \begin{tabular}{c|c|c|c|c}
        \hline\hline
        \textbf{Ratio} & \textbf{factorization} & \textbf{DG} &  \textbf{mDG} & \textbf{exp. avg.} \\ \hline
        $L_1/K$ & $0.1202(6)$& $0.1198$ & $0.1198$ &  $---$\\
        $L_2/K$ & $0.00097(1)$  & $0.001$ & $0.001$ & $---$\\
        $L_3/K$ & $---$ & $3.05\times 10^{-9}$ & $1.24\times 10^{-9}$ & $---$\\[+0.5ex]
        $\mathbf{L/K}$ & $\mathbf{0.1211(6)}$ & $\mathbf{0.1208}$ & $\mathbf{0.1208}$ & $\mathbf{0.1183(5)}$\\[+0.5ex]\hline 
        $M_1/K$ & $0.0212(2)$& $0.0212$ & $0.0212$ & $---$\\
        $M_2/K$ & $0.000170(3)$  & $0.000175$ & $0.000172$& $---$\\
        $M_3/K$ & $---$ & $5.10\times 10^{-10}$ & $2.07\times 10^{-10}$ & $---$\\
        $M_4/K$ & $---$  & $2.50\times 10^{-13}$ & $1.01\times 10^{-13}$& $---$\\
        $M_5/K$ & $---$ & $2.50\times 10^{-22}$ & $7.99\times 10^{-24}$& $---$\\[+0.5ex]
        $\mathbf{M/K}$ & $\mathbf{0.0214(2)}$ & $\mathbf{0.0213}$ & $\mathbf{0.0213}$ & $\mathbf{0.0199(7)}$\\\hline
         $\mathbf{M/L}$ & $\mathbf{0.177(1)}$ & $\mathbf{0.177}$ & $\mathbf{0.177}$ &$\mathbf{0.167(5)}$
        \\\hline\hline
    \end{tabular}}
    \caption{Theoretical capture ratios predicted in the factorized LO scheme compared to the non-factorized calculation for $\rho_{1,1}^{TD}$ fixed to the nuclear density reported in Ref.~\cite{Cadeddu:2025ueh}. In the last column, the theoretical predictions are compared to an averaging of the available experimental measurements~\cite{Bambynek,genz,Dougan,MANDUCHI-per-misure-sperimentali,CoGeNT,Collar:2023yew,edelweiss,CDMS,CONUS+}.}
    \label{tab:theoretical_fractions}
\end{table}

These can be compared to a reevaluation of the capture ratios obtained in the non-factorized scheme by fixing the phenomenological transition densities extracted in Ref.~\cite{Cadeddu:2025ueh} as possible solutions to the gallium anomaly. In Ref.~\cite{Cadeddu:2025ueh}, it was shown that a properly modeled transition density with at least one node can reduce significantly the neutrino capture cross section on $^{71}\text{Ga}$ while preserving the measured half-life of $^{71}\text{Ge}$.
To estimate the possible impact of such transition densities on the computation of capture ratios, we reevaluate them in the complete non-factorized scheme, in particular taking as an example the so-called DG (Double-Gaussian) and mDG (modified Double-Gaussian) parametrizations, with the former extending well beyond the nuclear radius, and the latter being much more compact.\\
The predictions of the capture ratios are reported in Tab.~\ref{tab:theoretical_fractions}, from which it can be noticed that the introduction of the full non-factorized scheme barely affects the prediction of the $L/K$ and $M/K$ capture ratios, but it allows one to predict the individual shell contributions also for the subdominant shells, such as the $L_3$ one, which are not included within the strict allowed-LO approximation. As expected, the transition density modifies the dominant contribution within the uncertainty independently of the specific parametrization considered for $\rho_{1,1}^{TD}(r)$. Instead, for the other subdominant shells the specific profile of the transition density can modify the prediction up to a factor of two. Unfortunately, these effects remain highly subdominant and are therefore not measurable within existing, and probably near-future, experiments.

\section{Comparison with existing measurements}
Our theoretical predictions need to be compared with the available experimental measurements. Therefore, in this work, we review existing data and combine them into an experimental average which describes the up-to-date experimental status.
\\
In the literature, a few measurements of $^{71}\text{Ge}$ electron capture ratios can be found, mainly focusing on the $K$, $L_1$ and $L_2$ captures, whose atomic de-excitation energies are $E_{K}\simeq 10.4\, \text{keV}$, $E_{L_1}\simeq 1.3\, \text{keV}$ and $E_{L_2}\simeq 1.1\, \text{keV}$, and thus are easier to measure. The $M$ shell instead is at $E_M\simeq 160\, \text{eV}$, closer to the threshold of modern rare-event detectors, such as CONUS+~\cite{CONUS+}.
\begin{table}[h]
\centering
    \resizebox{\columnwidth}{!}{
    \renewcommand{\arraystretch}{1.2}{}
    \begin{tabular}{l|c|c|c}
        \toprule\toprule
        \textbf{Reference} & \textbf{exp. $L/K$} & \textbf{exp. $M/K$} & \textbf{exp. $M/L$}\\
        \hline
        Konstantinov (1961) \cite{Bambynek}  & $0.13(2)$ & $---$ & $---$\\
        Dougan et al. (1961) \cite{Dougan}  & $0.116(5)$ & $0.019(9)$ & $0.16(8)$\\
        Manduchi et al. (1962) \cite{MANDUCHI-per-misure-sperimentali}       & $0.1187(8)$ & $0.017(1)$ & $0.14(1)$\\
        Genz et al. (1971) \cite{genz}           & $0.117(1)$ & $0.0190(4)$ & $0.162(3)$\\
        CoGeNT (2011) \cite{CoGeNT} & $0.116(4)$ & $---$ & $---$\\
        Collar et al. (2023) \cite{Collar:2023yew}         & $0.125(8)$ & $---$ & $---$\\
        EDELWEISS (2020) \cite{edelweiss}     & $0.110(8)$ & $0.017(3)$ & $0.16(2)$\\
        CDMS  (2016) \cite{CDMS}  & $---$ & $---$ &  $0.16(3)$\\
        CONUS+  (2026) \cite{CONUS+}        & $0.119(1)$ & $0.0214(4)$ & $0.180(4)$ \\
        \midrule
        Experimental average & $0.1183(5)$ & $0.0199(7)$ & $0.167(5)$ \\
        \bottomrule\bottomrule
    \end{tabular}}
        \caption{Experimental values of the $^{71}\text{Ge}$ capture ratios. 
         The values of $M/K$ are derived from the reported $L/K$ and $M/L$ ratios for those experiments sensitive to both the contributions except for CONUS+~\cite{CONUS+}, which instead reports $M/K$ instead of $M/L$.}
    \label{tab:LsuK valori sperimentali}
\end{table}

A compilation of measurements can be found in Ref.~\cite{Bambynek} and can be extended considering Refs.~\cite{genz,Dougan,MANDUCHI-per-misure-sperimentali,CoGeNT,Collar:2023yew,edelweiss,CDMS,CONUS+}, as reported in our Tab.~\ref{tab:LsuK valori sperimentali}. We report also the $M/L$ ratios determined by converting the corresponding $L/K$ and $M/K$ measurements for those experiments sensitive to both the contributions, and propagating the corresponding uncertainties.\\
As the data show some tensions among them, we account for such disagreements when evaluating the average experimental result, following the prescription of Ref.~\cite{average-procedure}. This procedure introduces a reduced chi-squared function to quantify the disagreement between the data, and in the case the latter exceeds unity, the resulting uncertainty is enlarged by the so-called Birge factor to account for such a tension. The resulting experimental averages are reported in the last row of Tab.~\ref{tab:LsuK valori sperimentali}.
\begin{figure}[h]
    \centering
    \includegraphics[width=\columnwidth]{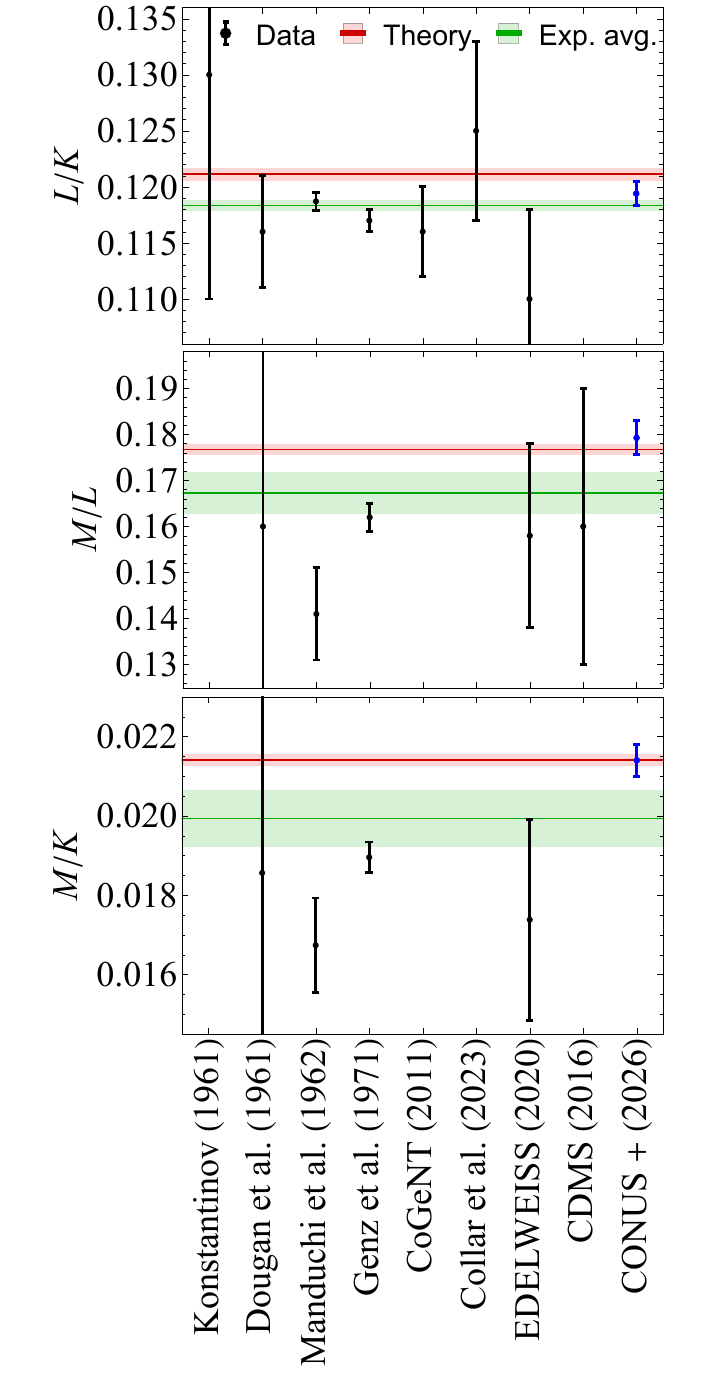}
    \caption{Summary of the available measurements of electron capture ratios of $^{71}\text{Ge}$ compared to our theoretical prediction (in red). The experimental average determined in this work is depicted by the green line with the corresponding $1\sigma$ uncertainty. We highlight with the blue color the latest experimental measurement by CONUS+~\cite{CONUS+}.}
    \label{fig:enter-label}
\end{figure}

In Fig.~\ref{fig:enter-label} we further compare the available measurements, and the corresponding experimental average, to our theoretical predictions. The existing tensions among the data are clearly visible, especially in the case of $M/L$ and $M/K$ capture ratios. In particular, the latest available measurement by CONUS+~\cite{CONUS+}, an experiment which looks for coherent elastic neutrino-nucleus scattering of reactor antineutrinos
on germanium nuclei, is depicted in Fig.~\ref{fig:enter-label} by the blue data point and shows an excellent agreement with our theoretical predictions while being in tension with older data, such as the Genz et al. result~\cite{genz}.\\
The disagreement between our theoretical prediction and the experimental average corresponds to a $3.6\sigma$ discrepancy in the case of $L/K$ capture ratios, while $\sim2\sigma$ for the $M/L$ and $M/K$ ones.
Additional atomic-model systematics, which are difficult to quantify reliably, may nevertheless be present. We therefore refer to the $3.6\sigma$ disagreement as the nominal discrepancy for $L/K$ ratios. As a robustness test, the inclusion of an additional relative theory uncertainty of 0.5\% (1\%) on the predicted L/K ratio would reduce the discrepancy to approximately $2.8\sigma$ ($1.9\sigma$).
\\
As the effect of the non-factorization is expected to be small, as discussed previously, it is unlikely that the observed disagreement could be solved in terms of nuclear structure effects. Taking the full difference between the factorized and non-factorized DG/mDG predictions as an additional model uncertainty reduces the nominal discrepancy from $3.6\sigma$ to approximately $3.3\sigma$.
Therefore, an explanation for such a tension is still to be found, and new data would help in clarifying the experimental picture, which as of now is rather complex.

\section{Conclusions}
\label{sec:conclusions}
In this work we developed a complete, non-factorized description of the electron capture process of $^{71}\text{Ge}$ to the ground state of $^{71}\text{Ga}$, based on unexpanded numerical DHFS electron wave functions coupled to a partial-wave expansion of the emitted neutrino. This approach allows the explicit calculation of the radial amplitudes for all relevant atomic shells ($K$, $L_{1-3}$ and $M_{1-5}$), going beyond the limitations of the traditional factorization of the leptonic contribution to the transition.\\
The analysis of the radial dependence of the leptonic kernels shows that the shells dominating the capture ($K$, $L_1$, $L_2$, $M_1$, $M_2$) share an essentially constant behavior, justifying the cancellation of the nuclear matrix element in the calculation of capture ratios. Therefore, the theoretical prediction of electron capture ratios of $^{71}\text{Ge}$ is largely insensitive to nuclear structure effects within the transition-density models considered, thus confirming the robustness of the factorized allowed-LO approximation for interpreting the currently available measurements. Conversely, subdominant shells such as $L_3$ display a different radial dependence, making the corresponding capture ratio potentially sensitive to the nuclear transition density, an effect, however, too small to be observable with current or near-future experimental sensitivity.\\
The comparison between predictions in the factorized and non-factorized schemes, performed using the phenomenological transition densities (DG and mDG) proposed as a possible solution to the gallium anomaly~\cite{Cadeddu:2025ueh}, confirms that the differences between the two approaches remain within uncertainties for the dominant $L/K$ and $M/K$ ratios.\\
We further compiled a new average of the experimental measurements available in the literature~\cite{Bambynek,genz,Dougan,MANDUCHI-per-misure-sperimentali,CoGeNT,Collar:2023yew,edelweiss,CDMS,CONUS+}, accounting for tensions among the different datasets through an error-inflation procedure~\cite{average-procedure}. Comparison with the theoretical predictions reveals a disagreement of about $3.6\sigma$ for the $L/K$ ratio and about $2\sigma$ for $M/K$ and $M/L$.
Despite these discrepancies with the experimental world averages, our predictions are consistent with the recent CONUS+ results~\cite{CONUS+}. In particular, CONUS+ clearly resolves the $^{71}\text{Ge}$ M-shell peak at $158.7\pm1.4\, \mathrm{eV}$, with the energy reconstruction validated down to the detection threshold, thus providing a particularly robust determination of the M-shell contribution. Since nuclear-structure effects are found to be insufficient to explain the discrepancies with the older data, new high-precision measurements will be essential to clarify the current experimental picture.

\bibliography{ref}

\begin{thebibliography}{30}%
\makeatletter
\providecommand \@ifxundefined [1]{%
 \@ifx{#1\undefined}
}%
\providecommand \@ifnum [1]{%
 \ifnum #1\expandafter \@firstoftwo
 \else \expandafter \@secondoftwo
 \fi
}%
\providecommand \@ifx [1]{%
 \ifx #1\expandafter \@firstoftwo
 \else \expandafter \@secondoftwo
 \fi
}%
\providecommand \natexlab [1]{#1}%
\providecommand \enquote  [1]{``#1''}%
\providecommand \bibnamefont  [1]{#1}%
\providecommand \bibfnamefont [1]{#1}%
\providecommand \citenamefont [1]{#1}%
\providecommand \href@noop [0]{\@secondoftwo}%
\providecommand \href [0]{\begingroup \@sanitize@url \@href}%
\providecommand \@href[1]{\@@startlink{#1}\@@href}%
\providecommand \@@href[1]{\endgroup#1\@@endlink}%
\providecommand \@sanitize@url [0]{\catcode `\\12\catcode `\$12\catcode
  `\&12\catcode `\#12\catcode `\^12\catcode `\_12\catcode `\%12\relax}%
\providecommand \@@startlink[1]{}%
\providecommand \@@endlink[0]{}%
\providecommand \url  [0]{\begingroup\@sanitize@url \@url }%
\providecommand \@url [1]{\endgroup\@href {#1}{\urlprefix }}%
\providecommand \urlprefix  [0]{URL }%
\providecommand \Eprint [0]{\href }%
\providecommand \doibase [0]{http://dx.doi.org/}%
\providecommand \selectlanguage [0]{\@gobble}%
\providecommand \bibinfo  [0]{\@secondoftwo}%
\providecommand \bibfield  [0]{\@secondoftwo}%
\providecommand \translation [1]{[#1]}%
\providecommand \BibitemOpen [0]{}%
\providecommand \bibitemStop [0]{}%
\providecommand \bibitemNoStop [0]{.\EOS\space}%
\providecommand \EOS [0]{\spacefactor3000\relax}%
\providecommand \BibitemShut  [1]{\csname bibitem#1\endcsname}%
\let\auto@bib@innerbib\@empty
\bibitem [{\citenamefont {Collar}\ and\ \citenamefont
  {Yoon}(2023)}]{Collar:2023yew}%
  \BibitemOpen
  \bibfield  {author} {\bibinfo {author} {\bibfnamefont {J.~I.}\ \bibnamefont
  {Collar}}\ and\ \bibinfo {author} {\bibfnamefont {S.~G.}\ \bibnamefont
  {Yoon}},\ }\href {\doibase 10.1103/PhysRevC.108.L021602} {\bibfield
  {journal} {\bibinfo  {journal} {Phys. Rev. C}\ }\textbf {\bibinfo {volume}
  {108}},\ \bibinfo {pages} {L021602} (\bibinfo {year} {2023})},\ \Eprint
  {http://arxiv.org/abs/2307.05353} {arXiv:2307.05353 [nucl-ex]} \BibitemShut
  {NoStop}%
\bibitem [{\citenamefont {Hampel}\ and\ \citenamefont
  {Remsberg}(1985)}]{PhysRevC.31.666}%
  \BibitemOpen
  \bibfield  {author} {\bibinfo {author} {\bibfnamefont {W.}~\bibnamefont
  {Hampel}}\ and\ \bibinfo {author} {\bibfnamefont {L.~P.}\ \bibnamefont
  {Remsberg}},\ }\href {\doibase 10.1103/PhysRevC.31.666} {\bibfield  {journal}
  {\bibinfo  {journal} {Phys. Rev. C}\ }\textbf {\bibinfo {volume} {31}},\
  \bibinfo {pages} {666} (\bibinfo {year} {1985})}\BibitemShut {NoStop}%
\bibitem [{\citenamefont {Derbin}\ \emph {et~al.}(2025)\citenamefont {Derbin},
  \citenamefont {Drachnev}, \citenamefont {Ivanov}, \citenamefont {Muratova},
  \citenamefont {Trushin}, \citenamefont {Unzhakov},\ and\ \citenamefont
  {Vorobyev}}]{newGelifetime}%
  \BibitemOpen
  \bibfield  {author} {\bibinfo {author} {\bibfnamefont {A.}~\bibnamefont
  {Derbin}}, \bibinfo {author} {\bibfnamefont {I.}~\bibnamefont {Drachnev}},
  \bibinfo {author} {\bibfnamefont {D.}~\bibnamefont {Ivanov}}, \bibinfo
  {author} {\bibfnamefont {V.}~\bibnamefont {Muratova}}, \bibinfo {author}
  {\bibfnamefont {M.}~\bibnamefont {Trushin}}, \bibinfo {author} {\bibfnamefont
  {E.}~\bibnamefont {Unzhakov}}, \ and\ \bibinfo {author} {\bibfnamefont
  {A.}~\bibnamefont {Vorobyev}},\ }\href {\doibase 10.1134/S1063778825700152}
  {\bibfield  {journal} {\bibinfo  {journal} {Physics of Atomic Nuclei}\
  }\textbf {\bibinfo {volume} {88}},\ \bibinfo {pages} {52} (\bibinfo {year}
  {2025})}\BibitemShut {NoStop}%
\bibitem [{\citenamefont {Norman}\ \emph {et~al.}(2024)\citenamefont {Norman},
  \citenamefont {Drobizhev}, \citenamefont {Gharibyan}, \citenamefont
  {Gregorich}, \citenamefont {Kolomensky}, \citenamefont {Sammis},
  \citenamefont {Scielzo}, \citenamefont {Shusterman},\ and\ \citenamefont
  {Thomas}}]{PhysRevC.109.055501}%
  \BibitemOpen
  \bibfield  {author} {\bibinfo {author} {\bibfnamefont {E.~B.}\ \bibnamefont
  {Norman}}, \bibinfo {author} {\bibfnamefont {A.}~\bibnamefont {Drobizhev}},
  \bibinfo {author} {\bibfnamefont {N.}~\bibnamefont {Gharibyan}}, \bibinfo
  {author} {\bibfnamefont {K.~E.}\ \bibnamefont {Gregorich}}, \bibinfo {author}
  {\bibfnamefont {Y.~G.}\ \bibnamefont {Kolomensky}}, \bibinfo {author}
  {\bibfnamefont {B.~N.}\ \bibnamefont {Sammis}}, \bibinfo {author}
  {\bibfnamefont {N.~D.}\ \bibnamefont {Scielzo}}, \bibinfo {author}
  {\bibfnamefont {J.~A.}\ \bibnamefont {Shusterman}}, \ and\ \bibinfo {author}
  {\bibfnamefont {K.~J.}\ \bibnamefont {Thomas}},\ }\href {\doibase
  10.1103/PhysRevC.109.055501} {\bibfield  {journal} {\bibinfo  {journal}
  {Phys. Rev. C}\ }\textbf {\bibinfo {volume} {109}},\ \bibinfo {pages}
  {055501} (\bibinfo {year} {2024})}\BibitemShut {NoStop}%
\bibitem [{\citenamefont {Singh}\ and\ \citenamefont {Chen}(2023)}]{Qvalue}%
  \BibitemOpen
  \bibfield  {author} {\bibinfo {author} {\bibfnamefont {B.}~\bibnamefont
  {Singh}}\ and\ \bibinfo {author} {\bibfnamefont {J.}~\bibnamefont {Chen}},\
  }\href {\doibase https://doi.org/10.1016/j.nds.2023.02.001} {\bibfield
  {journal} {\bibinfo  {journal} {Nuclear Data Sheets}\ }\textbf {\bibinfo
  {volume} {188}},\ \bibinfo {pages} {1} (\bibinfo {year} {2023})}\BibitemShut
  {NoStop}%
\bibitem [{\citenamefont {Bambynek}\ \emph {et~al.}(1977)\citenamefont
  {Bambynek}, \citenamefont {Behrens}, \citenamefont {Chen}, \citenamefont
  {Crasemann}, \citenamefont {Fitzpatrick}, \citenamefont {Ledingham},
  \citenamefont {Genz}, \citenamefont {Mutterer},\ and\ \citenamefont
  {Intemann}}]{Bambynek}%
  \BibitemOpen
  \bibfield  {author} {\bibinfo {author} {\bibfnamefont {W.}~\bibnamefont
  {Bambynek}}, \bibinfo {author} {\bibfnamefont {H.}~\bibnamefont {Behrens}},
  \bibinfo {author} {\bibfnamefont {M.~H.}\ \bibnamefont {Chen}}, \bibinfo
  {author} {\bibfnamefont {B.}~\bibnamefont {Crasemann}}, \bibinfo {author}
  {\bibfnamefont {M.~L.}\ \bibnamefont {Fitzpatrick}}, \bibinfo {author}
  {\bibfnamefont {K.~W.~D.}\ \bibnamefont {Ledingham}}, \bibinfo {author}
  {\bibfnamefont {H.}~\bibnamefont {Genz}}, \bibinfo {author} {\bibfnamefont
  {M.}~\bibnamefont {Mutterer}}, \ and\ \bibinfo {author} {\bibfnamefont
  {R.~L.}\ \bibnamefont {Intemann}},\ }\href {\doibase
  10.1103/RevModPhys.49.77} {\bibfield  {journal} {\bibinfo  {journal} {Rev.
  Mod. Phys.}\ }\textbf {\bibinfo {volume} {49}},\ \bibinfo {pages} {77}
  (\bibinfo {year} {1977})}\BibitemShut {NoStop}%
\bibitem [{\citenamefont {Cadeddu}\ \emph {et~al.}(2025)\citenamefont
  {Cadeddu}, \citenamefont {Cargioli}, \citenamefont {Dordei}, \citenamefont
  {Ferro}, \citenamefont {Giunti},\ and\ \citenamefont
  {Pitzalis}}]{Cadeddu:2025ueh}%
  \BibitemOpen
  \bibfield  {author} {\bibinfo {author} {\bibfnamefont {M.}~\bibnamefont
  {Cadeddu}}, \bibinfo {author} {\bibfnamefont {N.}~\bibnamefont {Cargioli}},
  \bibinfo {author} {\bibfnamefont {F.}~\bibnamefont {Dordei}}, \bibinfo
  {author} {\bibfnamefont {L.}~\bibnamefont {Ferro}}, \bibinfo {author}
  {\bibfnamefont {C.}~\bibnamefont {Giunti}}, \ and\ \bibinfo {author}
  {\bibfnamefont {M.}~\bibnamefont {Pitzalis}},\ }\href@noop {} {\  (\bibinfo
  {year} {2025})},\ \Eprint {http://arxiv.org/abs/2512.20560} {arXiv:2512.20560
  [hep-ph]} \BibitemShut {NoStop}%
\bibitem [{\citenamefont {Salvat}\ and\ \citenamefont
  {Fern\'andez-Varea}(2019)}]{radial}%
  \BibitemOpen
  \bibfield  {author} {\bibinfo {author} {\bibfnamefont {F.}~\bibnamefont
  {Salvat}}\ and\ \bibinfo {author} {\bibfnamefont {J.~M.}\ \bibnamefont
  {Fern\'andez-Varea}},\ }\href {\doibase
  https://doi.org/10.1016/j.cpc.2019.02.011} {\bibfield  {journal} {\bibinfo
  {journal} {Computer Physics Communications}\ }\textbf {\bibinfo {volume}
  {240}},\ \bibinfo {pages} {165} (\bibinfo {year} {2019})}\BibitemShut
  {NoStop}%
\bibitem [{\citenamefont {Ravli\ifmmode~\acute{c}\else \'{c}\fi{}}\ \emph
  {et~al.}(2025)\citenamefont {Ravli\ifmmode~\acute{c}\else \'{c}\fi{}},
  \citenamefont {Schwerdtfeger},\ and\ \citenamefont {Nazarewicz}}]{n6d8-cmcg}%
  \BibitemOpen
  \bibfield  {author} {\bibinfo {author} {\bibfnamefont {A.}~\bibnamefont
  {Ravli\ifmmode~\acute{c}\else \'{c}\fi{}}}, \bibinfo {author} {\bibfnamefont
  {P.}~\bibnamefont {Schwerdtfeger}}, \ and\ \bibinfo {author} {\bibfnamefont
  {W.}~\bibnamefont {Nazarewicz}},\ }\href {\doibase 10.1103/n6d8-cmcg}
  {\bibfield  {journal} {\bibinfo  {journal} {Phys. Rev. C}\ }\textbf {\bibinfo
  {volume} {111}},\ \bibinfo {pages} {064323} (\bibinfo {year}
  {2025})}\BibitemShut {NoStop}%
\bibitem [{\citenamefont {Navas}\ \emph {et~al.}(2024)\citenamefont {Navas}
  \emph {et~al.}}]{ParticleDataGroup:2024cfk}%
  \BibitemOpen
  \bibfield  {author} {\bibinfo {author} {\bibfnamefont {S.}~\bibnamefont
  {Navas}} \emph {et~al.} (\bibinfo {collaboration} {Particle Data Group}),\
  }\href {\doibase 10.1103/PhysRevD.110.030001} {\bibfield  {journal} {\bibinfo
   {journal} {Phys. Rev. D}\ }\textbf {\bibinfo {volume} {110}},\ \bibinfo
  {pages} {030001} (\bibinfo {year} {2024})}\BibitemShut {NoStop}%
\bibitem [{\citenamefont {M\"arkisch}\ \emph {et~al.}(2019)\citenamefont
  {M\"arkisch}, \citenamefont {Mest}, \citenamefont {Saul}, \citenamefont
  {Wang}, \citenamefont {Abele}, \citenamefont {Dubbers}, \citenamefont
  {Klopf}, \citenamefont {Petoukhov}, \citenamefont {Roick}, \citenamefont
  {Soldner},\ and\ \citenamefont {Werder}}]{PERKEO3}%
  \BibitemOpen
  \bibfield  {author} {\bibinfo {author} {\bibfnamefont {B.}~\bibnamefont
  {M\"arkisch}}, \bibinfo {author} {\bibfnamefont {H.}~\bibnamefont {Mest}},
  \bibinfo {author} {\bibfnamefont {H.}~\bibnamefont {Saul}}, \bibinfo {author}
  {\bibfnamefont {X.}~\bibnamefont {Wang}}, \bibinfo {author} {\bibfnamefont
  {H.}~\bibnamefont {Abele}}, \bibinfo {author} {\bibfnamefont
  {D.}~\bibnamefont {Dubbers}}, \bibinfo {author} {\bibfnamefont
  {M.}~\bibnamefont {Klopf}}, \bibinfo {author} {\bibfnamefont
  {A.}~\bibnamefont {Petoukhov}}, \bibinfo {author} {\bibfnamefont
  {C.}~\bibnamefont {Roick}}, \bibinfo {author} {\bibfnamefont
  {T.}~\bibnamefont {Soldner}}, \ and\ \bibinfo {author} {\bibfnamefont
  {D.}~\bibnamefont {Werder}},\ }\href {\doibase
  10.1103/PhysRevLett.122.242501} {\bibfield  {journal} {\bibinfo  {journal}
  {Phys. Rev. Lett.}\ }\textbf {\bibinfo {volume} {122}},\ \bibinfo {pages}
  {242501} (\bibinfo {year} {2019})}\BibitemShut {NoStop}%
\bibitem [{\citenamefont {Elliott}\ \emph {et~al.}(2023)\citenamefont
  {Elliott}, \citenamefont {Gavrin}, \citenamefont {Haxton}, \citenamefont
  {Ibragimova},\ and\ \citenamefont {Rule}}]{Elliott:2023xkb}%
  \BibitemOpen
  \bibfield  {author} {\bibinfo {author} {\bibfnamefont {S.~R.}\ \bibnamefont
  {Elliott}}, \bibinfo {author} {\bibfnamefont {V.~N.}\ \bibnamefont {Gavrin}},
  \bibinfo {author} {\bibfnamefont {W.~C.}\ \bibnamefont {Haxton}}, \bibinfo
  {author} {\bibfnamefont {T.~V.}\ \bibnamefont {Ibragimova}}, \ and\ \bibinfo
  {author} {\bibfnamefont {E.~J.}\ \bibnamefont {Rule}},\ }\href {\doibase
  10.1103/PhysRevC.108.035502} {\bibfield  {journal} {\bibinfo  {journal}
  {Phys. Rev. C}\ }\textbf {\bibinfo {volume} {108}},\ \bibinfo {pages}
  {035502} (\bibinfo {year} {2023})},\ \Eprint
  {http://arxiv.org/abs/2303.13623} {arXiv:2303.13623 [nucl-th]} \BibitemShut
  {NoStop}%
\bibitem [{\citenamefont {Bahcall}(1962)}]{PhysRevLett.9.500}%
  \BibitemOpen
  \bibfield  {author} {\bibinfo {author} {\bibfnamefont {J.~N.}\ \bibnamefont
  {Bahcall}},\ }\href {\doibase 10.1103/PhysRevLett.9.500} {\bibfield
  {journal} {\bibinfo  {journal} {Phys. Rev. Lett.}\ }\textbf {\bibinfo
  {volume} {9}},\ \bibinfo {pages} {500} (\bibinfo {year} {1962})}\BibitemShut
  {NoStop}%
\bibitem [{\citenamefont {Vatai}(1970)}]{Vatai:1970ldf}%
  \BibitemOpen
  \bibfield  {author} {\bibinfo {author} {\bibfnamefont {E.}~\bibnamefont
  {Vatai}},\ }\href {\doibase 10.1016/0375-9474(70)90250-2} {\bibfield
  {journal} {\bibinfo  {journal} {Nucl. Phys. A}\ }\textbf {\bibinfo {volume}
  {156}},\ \bibinfo {pages} {541} (\bibinfo {year} {1970})}\BibitemShut
  {NoStop}%
\bibitem [{\citenamefont {Bahcall}(1963{\natexlab{a}})}]{PhysRev.131.1756}%
  \BibitemOpen
  \bibfield  {author} {\bibinfo {author} {\bibfnamefont {J.~N.}\ \bibnamefont
  {Bahcall}},\ }\href {\doibase 10.1103/PhysRev.131.1756} {\bibfield  {journal}
  {\bibinfo  {journal} {Phys. Rev.}\ }\textbf {\bibinfo {volume} {131}},\
  \bibinfo {pages} {1756} (\bibinfo {year} {1963}{\natexlab{a}})}\BibitemShut
  {NoStop}%
\bibitem [{\citenamefont {Bahcall}(1963{\natexlab{b}})}]{Bahcall:1963zza}%
  \BibitemOpen
  \bibfield  {author} {\bibinfo {author} {\bibfnamefont {J.~N.}\ \bibnamefont
  {Bahcall}},\ }\href {\doibase 10.1103/PhysRev.132.362} {\bibfield  {journal}
  {\bibinfo  {journal} {Phys. Rev.}\ }\textbf {\bibinfo {volume} {132}},\
  \bibinfo {pages} {362} (\bibinfo {year} {1963}{\natexlab{b}})}\BibitemShut
  {NoStop}%
\bibitem [{\citenamefont {Bahcall}(1965)}]{BAHCALL1965267}%
  \BibitemOpen
  \bibfield  {author} {\bibinfo {author} {\bibfnamefont {J.~N.}\ \bibnamefont
  {Bahcall}},\ }\href {\doibase https://doi.org/10.1016/0029-5582(65)90717-0}
  {\bibfield  {journal} {\bibinfo  {journal} {Nuclear Physics}\ }\textbf
  {\bibinfo {volume} {71}},\ \bibinfo {pages} {267} (\bibinfo {year}
  {1965})}\BibitemShut {NoStop}%
\bibitem [{\citenamefont {Mougeot}(2018)}]{Mougeot:2018whu}%
  \BibitemOpen
  \bibfield  {author} {\bibinfo {author} {\bibfnamefont {X.}~\bibnamefont
  {Mougeot}},\ }\href {\doibase 10.1016/j.apradiso.2017.07.027} {\bibfield
  {journal} {\bibinfo  {journal} {Appl. Radiat. Isot.}\ }\textbf {\bibinfo
  {volume} {134}},\ \bibinfo {pages} {225} (\bibinfo {year}
  {2018})}\BibitemShut {NoStop}%
\bibitem [{\citenamefont {Sevestrean}\ \emph {et~al.}(2023)\citenamefont
  {Sevestrean}, \citenamefont {Ni{\c{t}}escu}, \citenamefont {Ghinescu},\ and\
  \citenamefont {Stoica}}]{Sevestrean:2023qzm}%
  \BibitemOpen
  \bibfield  {author} {\bibinfo {author} {\bibfnamefont {V.~A.}\ \bibnamefont
  {Sevestrean}}, \bibinfo {author} {\bibfnamefont {O.}~\bibnamefont
  {Ni{\c{t}}escu}}, \bibinfo {author} {\bibfnamefont {S.}~\bibnamefont
  {Ghinescu}}, \ and\ \bibinfo {author} {\bibfnamefont {S.}~\bibnamefont
  {Stoica}},\ }\href {\doibase 10.1103/PhysRevA.108.012810} {\bibfield
  {journal} {\bibinfo  {journal} {Phys. Rev. A}\ }\textbf {\bibinfo {volume}
  {108}},\ \bibinfo {pages} {012810} (\bibinfo {year} {2023})},\ \Eprint
  {http://arxiv.org/abs/2304.10373} {arXiv:2304.10373 [nucl-th]} \BibitemShut
  {NoStop}%
\bibitem [{\citenamefont {Cadeddu}\ \emph {et~al.}(2026)\citenamefont
  {Cadeddu}, \citenamefont {Cargioli}, \citenamefont {Carotenuto},
  \citenamefont {Dordei}, \citenamefont {Ferro},\ and\ \citenamefont
  {Giunti}}]{Cadeddu:2025pue}%
  \BibitemOpen
  \bibfield  {author} {\bibinfo {author} {\bibfnamefont {M.}~\bibnamefont
  {Cadeddu}}, \bibinfo {author} {\bibfnamefont {N.}~\bibnamefont {Cargioli}},
  \bibinfo {author} {\bibfnamefont {G.}~\bibnamefont {Carotenuto}}, \bibinfo
  {author} {\bibfnamefont {F.}~\bibnamefont {Dordei}}, \bibinfo {author}
  {\bibfnamefont {L.}~\bibnamefont {Ferro}}, \ and\ \bibinfo {author}
  {\bibfnamefont {C.}~\bibnamefont {Giunti}},\ }\href {\doibase
  10.1103/pz8g-zz1b} {\bibfield  {journal} {\bibinfo  {journal} {Phys. Rev. D}\
  }\textbf {\bibinfo {volume} {113}},\ \bibinfo {pages} {033006} (\bibinfo
  {year} {2026})},\ \Eprint {http://arxiv.org/abs/2507.13103} {arXiv:2507.13103
  [hep-ph]} \BibitemShut {NoStop}%
\bibitem [{\citenamefont {King}\ \emph {et~al.}(2020)\citenamefont {King},
  \citenamefont {Andreoli}, \citenamefont {Pastore},\ and\ \citenamefont
  {Piarulli}}]{saori-pastore}%
  \BibitemOpen
  \bibfield  {author} {\bibinfo {author} {\bibfnamefont {G.~B.}\ \bibnamefont
  {King}}, \bibinfo {author} {\bibfnamefont {L.}~\bibnamefont {Andreoli}},
  \bibinfo {author} {\bibfnamefont {S.}~\bibnamefont {Pastore}}, \ and\
  \bibinfo {author} {\bibfnamefont {M.}~\bibnamefont {Piarulli}},\ }\href
  {\doibase 10.3389/fphy.2020.00363} {\bibfield  {journal} {\bibinfo  {journal}
  {Frontiers in Physics}\ }\textbf {\bibinfo {volume} {Volume 8 - 2020}}
  (\bibinfo {year} {2020}),\ 10.3389/fphy.2020.00363}\BibitemShut {NoStop}%
\bibitem [{\citenamefont {Sch\"onfeld}(1998)}]{SCHONFELD19981353}%
  \BibitemOpen
  \bibfield  {author} {\bibinfo {author} {\bibfnamefont {E.}~\bibnamefont
  {Sch\"onfeld}},\ }\href {\doibase
  https://doi.org/10.1016/S0969-8043(97)10073-2} {\bibfield  {journal}
  {\bibinfo  {journal} {Applied Radiation and Isotopes}\ }\textbf {\bibinfo
  {volume} {49}},\ \bibinfo {pages} {1353} (\bibinfo {year}
  {1998})}\BibitemShut {NoStop}%
\bibitem [{\citenamefont {Genz}\ \emph {et~al.}(1971)\citenamefont {Genz},
  \citenamefont {Renier}, \citenamefont {Pengra},\ and\ \citenamefont
  {Fink}}]{genz}%
  \BibitemOpen
  \bibfield  {author} {\bibinfo {author} {\bibfnamefont {H.}~\bibnamefont
  {Genz}}, \bibinfo {author} {\bibfnamefont {J.~P.}\ \bibnamefont {Renier}},
  \bibinfo {author} {\bibfnamefont {J.~G.}\ \bibnamefont {Pengra}}, \ and\
  \bibinfo {author} {\bibfnamefont {R.~W.}\ \bibnamefont {Fink}},\ }\href
  {\doibase 10.1103/PhysRevC.3.172} {\bibfield  {journal} {\bibinfo  {journal}
  {Phys. Rev. C}\ }\textbf {\bibinfo {volume} {3}},\ \bibinfo {pages} {172}
  (\bibinfo {year} {1971})}\BibitemShut {NoStop}%
\bibitem [{\citenamefont {Dougan}()}]{Dougan}%
  \BibitemOpen
  \bibfield  {author} {\bibinfo {author} {\bibfnamefont {P.~W.}\ \bibnamefont
  {Dougan}},\ }\href@noop {} {\enquote {\bibinfo {title} {Proportional counter
  studies of orbital electron capture ratios},}\ }\bibinfo {howpublished}
  {Ph.D. thesis, University of Glasgow (1961), unpublished}\BibitemShut
  {NoStop}%
\bibitem [{\citenamefont {Manduchi}\ and\ \citenamefont
  {Zannoni}(1962)}]{MANDUCHI-per-misure-sperimentali}%
  \BibitemOpen
  \bibfield  {author} {\bibinfo {author} {\bibfnamefont {C.}~\bibnamefont
  {Manduchi}}\ and\ \bibinfo {author} {\bibfnamefont {G.}~\bibnamefont
  {Zannoni}},\ }\href {\doibase https://doi.org/10.1016/0029-5582(62)90472-8}
  {\bibfield  {journal} {\bibinfo  {journal} {Nuclear Physics}\ }\textbf
  {\bibinfo {volume} {36}},\ \bibinfo {pages} {497} (\bibinfo {year}
  {1962})}\BibitemShut {NoStop}%
\bibitem [{\citenamefont {Aalseth}\ \emph {et~al.}(2011)\citenamefont {Aalseth}
  \emph {et~al.}}]{CoGeNT}%
  \BibitemOpen
  \bibfield  {author} {\bibinfo {author} {\bibfnamefont {C.~E.}\ \bibnamefont
  {Aalseth}} \emph {et~al.} (\bibinfo {collaboration} {CoGeNT Collaboration}),\
  }\href {\doibase 10.1103/PhysRevLett.106.131301} {\bibfield  {journal}
  {\bibinfo  {journal} {Phys. Rev. Lett.}\ }\textbf {\bibinfo {volume} {106}},\
  \bibinfo {pages} {131301} (\bibinfo {year} {2011})}\BibitemShut {NoStop}%
\bibitem [{\citenamefont {Arnaud}\ \emph {et~al.}(2020)\citenamefont {Arnaud}
  \emph {et~al.}}]{edelweiss}%
  \BibitemOpen
  \bibfield  {author} {\bibinfo {author} {\bibfnamefont {Q.}~\bibnamefont
  {Arnaud}} \emph {et~al.} (\bibinfo {collaboration} {EDELWEISS
  Collaboration}),\ }\href {\doibase 10.1103/PhysRevLett.125.141301} {\bibfield
   {journal} {\bibinfo  {journal} {Phys. Rev. Lett.}\ }\textbf {\bibinfo
  {volume} {125}},\ \bibinfo {pages} {141301} (\bibinfo {year}
  {2020})}\BibitemShut {NoStop}%
\bibitem [{\citenamefont {Agnese}\ \emph {et~al.}(2016)\citenamefont {Agnese}
  \emph {et~al.}}]{CDMS}%
  \BibitemOpen
  \bibfield  {author} {\bibinfo {author} {\bibfnamefont {R.}~\bibnamefont
  {Agnese}} \emph {et~al.} (\bibinfo {collaboration} {SuperCDMS
  Collaboration}),\ }\href {\doibase 10.1103/PhysRevLett.116.071301} {\bibfield
   {journal} {\bibinfo  {journal} {Phys. Rev. Lett.}\ }\textbf {\bibinfo
  {volume} {116}},\ \bibinfo {pages} {071301} (\bibinfo {year}
  {2016})}\BibitemShut {NoStop}%
\bibitem [{\citenamefont {S{\'a}nchez~Garc{\'\i}a}\ \emph
  {et~al.}(2026)\citenamefont {S{\'a}nchez~Garc{\'\i}a} \emph
  {et~al.}}]{CONUS+}%
  \BibitemOpen
  \bibfield  {author} {\bibinfo {author} {\bibfnamefont {E.}~\bibnamefont
  {S{\'a}nchez~Garc{\'\i}a}} \emph {et~al.},\ }\href@noop {} {\  (\bibinfo
  {year} {2026})},\ \Eprint {http://arxiv.org/abs/2604.25748} {arXiv:2604.25748
  [hep-ex]} \BibitemShut {NoStop}%
\bibitem [{\citenamefont {Trassinelli}\ and\ \citenamefont
  {Maxton}(2026)}]{average-procedure}%
  \BibitemOpen
  \bibfield  {author} {\bibinfo {author} {\bibfnamefont {M.}~\bibnamefont
  {Trassinelli}}\ and\ \bibinfo {author} {\bibfnamefont {M.}~\bibnamefont
  {Maxton}},\ }\href {\doibase 10.1088/1681-7575/ae4237} {\bibfield  {journal}
  {\bibinfo  {journal} {Metrologia}\ }\textbf {\bibinfo {volume} {63}},\
  \bibinfo {pages} {015009} (\bibinfo {year} {2026})},\ \Eprint
  {http://arxiv.org/abs/2406.08293} {arXiv:2406.08293 [physics.data-an]}
  \BibitemShut {NoStop}%
\end{thebibliography}%

\end{document}